\documentclass[twocolumn,preprintnumbers,amsmath,amssymb,superscriptaddress]{revtex4}
\usepackage{graphicx}
\usepackage{dcolumn}
\usepackage{bm}
\usepackage{soul}
\usepackage{color}
\usepackage{epstopdf}
\usepackage[version=3]{mhchem}
\usepackage{lipsum}
\usepackage[outercaption]{sidecap}
\usepackage{floatrow}
\usepackage{hyperref}
\usepackage{appendix}

\begin{document}

\title{Modification of the Statistical Moment Method for the High-Pressure Melting Curve by the Inclusion of Thermal Vacancies}
\author{Tran Dinh Cuong}
\email{cuong.trandinh@phenikaa-uni.edu.vn}
\affiliation{Phenikaa Institute for Advanced Study (PIAS), Phenikaa University, Hanoi 12116, Vietnam}
\author{Anh D. Phan}
\email{anh.phanduc@phenikaa-uni.edu.vn}
\affiliation{Phenikaa Institute for Advanced Study (PIAS), Phenikaa University, Hanoi 12116, Vietnam}
\affiliation{Faculty of Information Technology, Materials Science and Engineering, Artificial Intelligence Laboratory,  Phenikaa University, Hanoi 12116, Vietnam}
\date{\today}

\date{\today}% It is always \today, today,

\begin{abstract}
Melting behaviors of defective crystals under extreme conditions are theoretically investigated using the statistical moment method. In our theoretical model, heating processes cause missing atoms or vacancies in crystal structures via dislocating them from their equilibrium positions. The coordination number of some atoms is assumed to be removed by one unit and the defect depends on temperature and external pressure. We formulate analytical expressions to directly connect the equilibrium vacancy concentration, the elastic modulus, and the melting temperature. Numerical calculations are carried out for six transition metals including Cu, Ag, Au, Mo, Ta, and W up to 400 GPa. The obtained results show that vacancies strongly drive the melting transition. Ignoring the vacancy formation leads to incorrect predictions of the melting point. The good agreement between our numerical data and prior experimental works validates the accuracy of our approach.

\end{abstract}

\maketitle

%\tableofcontents

\section{INTRODUCTION}
Melting behaviors of crystals at ultra-high pressures have been extensively studied for decades due to their importance in condensed-matter physics \cite{1}, geophysics \cite{2}, and technology domains \cite{3}. Specifically, understandings of the melting of metallic materials can be used to explain internal dynamics and thermal profiles of planets in the solar systems \cite{4}. In addition, the melting process is a key factor for developing and optimizing modern manufacturing techniques, particularly as 3D printing \cite{5}.

There are four main techniques to explore the melting transition including: (1) laser-heated diamond anvil cell (LH DAC) measurements \cite{6}, (2) dynamic shock-wave experiments \cite{7}, (3) first-principle calculations \cite{8}, and (4) molecular dynamics simulations \cite{9}. However, a large disagreement among these methods has arisen in a variety of transition metals having a body-centered cubic (BCC) structure. For example, the melting temperature of Mo, Ta, and W reported by LH DAC \cite{10} grows slowly with compression and saturates at 100 GPa. In contrast, both dynamic shock-wave induced melting experiments and simulations suggest a strong increase of melting temperature up to 400 GPa \cite{11}. The deviation between LH DAC and other approaches in the case of Mo, Ta, and W can be up to thousands of Kelvin, which is far beyond a numerical tolerance \cite{10,11}. Although simulations provide calculations of the very high-pressure melting curve, most simulation approaches require a great computational cost and are very time-consuming \cite{12}. Particularly, the simulation complexity increases dramatically with compression and size of simulated systems \cite{13}. Moreover, it is difficult to access the melting boundary of multicomponent systems \cite{14}. Despite much effort over past few decades to overcome these challenges, a common consensus has not been achieved.

Recently, the statistical moment method (SMM) has been developed to investigate the melting properties of crystalline materials \cite{15}. The approach is based on analyzing the thermodynamic instability of the solid phase \cite{16} and the melting-mechanical property relations \cite{17} to establish an analytical expression for the melting curve without facing the complex dualities of liquids. In Ref.\cite{18}, for the first time, authors have found a simple correlation between the equilibrium vacancy concentration and the melting temperature via SMM calculations. They consider effects of the vacancy formation due to thermal excitations \cite{18}. The appearance of vacancies breaks down the lattice order, enhances molecular mobility, reduces the refractory ability, and promotes the melting transition of crystals \cite{19,20,21}. This method can help us to effectively capture the melting mechanism of metals \cite{18,22} and alloys \cite{23}. However, numerical calculations for vacancy effects are valid in a range of 0 to 100 GPa \cite{18,22,23}. At ultra-high pressures, the melting temperature predicted by the extended version of SMM method varies non-monotonically with compression, and this variation disagrees with experiments \cite{18,22,23}.  

In this paper, we develop the above theoretical model to resolve the mentioned limitations. Our modified SMM analysis directly links the melting point of perfect crystals to defective counterparts. To validate our approach, we carry out numerical calculations for Cu, Ag, and Au, where all of the existing data are quantitatively consistent \cite{24,25,26}. Then, we investigate the high-pressure melting curve of Mo, Ta, and W to provide an insight into the melting process. Mo, Ta, and W are of technological importance, but the information about their solid-liquid boundaries is alarmingly self-conflicting \cite{27,28,29}.    

\section{THEORETICAL BACKGROUND}
In this section, we briefly summarize the original SMM method for determining the melting properties of perfect and imperfect crystals. Then, we propose a new extension to appropriately take into account vacancy effects.
\subsection{Ideal Model}
In the SMM, for perfect crystals with cubic symmetry, where all atoms are in identical environments, the continuous periodic structure is characterized by the cohesive energy $u_0$, an effective spring hardness for each atom $k$, and anharmonic parameters $\gamma_1$, $\gamma_2$ and $\gamma$ \cite{30,31}. These quantities are defined by 
\begin{eqnarray}
u_0&=&\sum_i\varphi_{i0},\quad k=\frac{1}{2}\left(\frac{\partial^2u_0}{\partial u_x^2}\right)_{eq}=m\omega^2\nonumber\\
\gamma_1&=&\frac{1}{48}\left(\frac{\partial^4u_0}{\partial u_x^4}\right)_{eq},\quad\gamma_2=\frac{1}{8}\left(\frac{\partial^4u_0}{\partial u_x^2\partial u_y^2}\right)_{eq},\nonumber\\
\gamma&=&4(\gamma_1+\gamma_2),
\label{eq:1}
\end{eqnarray}
where $\varphi_{i0}$ is the pair interatomic potential between $i$-th atom and the central $0$-th atom, $m$ is the atomic mass, $\omega$ is the Einstein frequency, $u_x$ and $u_y$ are atomic displacements in $x$ and $y$ directions, respectively \cite{30,31}. From these, one can employ the equation of state \cite{32} to consider effects of external conditions on atomic arrangement, which is 
\begin{eqnarray}
Pv=-a\left(\frac{1}{6}\frac{\partial u_0}{\partial a}+\theta X\frac{1}{2k}\frac{\partial k}{\partial a}\right),
\label{eq:2}
\end{eqnarray}
where $P$ is the static pressure, $v$ is the atomic volume, $a$ is the nearest neighbor distance between lattice nodes, $\theta=k_BT$ is the thermal energy, $k_B$ is the Boltzmann constant, $T$ is temperature, $X=x\coth{x}$ is a normalized quantity, $x=\hbar\omega/2\theta$ denotes the quantization of energy, and $\hbar$ is the reduced Planck constant. At $T=0$, Eq. (\ref{eq:2}) is rewritten by
\begin{eqnarray}
Pv=-a\left(\frac{1}{6}\frac{\partial u_0}{\partial a}+\frac{\hbar \omega}{4k}\frac{\partial k}{\partial a}\right).
\label{eq:3}
\end{eqnarray}
Solving Eq. (\ref{eq:3}) gives the value of $a\equiv a(P,0)$ at 0 K. When $T>0$, we can compute $a(P,T)$ as
\begin{eqnarray}
a(P,T)=a(P,0)+\sqrt{\frac{2\gamma\theta^2}{3k^3}A},
\label{eq:4}
\end{eqnarray}
where the second term is the lattice expansion upon heating, and $A$ was analytically presented in Ref.\cite{32} by using a force balance condition for the central $0-$th atom.

The absolute stability limit of the crystalline state is \cite{16}
\begin{eqnarray}
\left(\cfrac{\partial P}{\partial v}\right)_{T}=0,
\label{eq:5}
\end{eqnarray}
Note that Eq. (\ref{eq:5}) is a theoretical hypothesis because it is equivalent to an infinity value of the isothermal compressibility. However, in the original SMM analysis \cite{15,16}, the critical temperature $T_s$ satisfying Eq.(\ref{eq:5}) is expected to be close to the melting temperature $T_m$. Thus, we can use $T_s$ to estimate $T_m$ via a simple correction and the absence of the liquid phase \cite{15,16}. Combining Eq. (\ref{eq:2}) and (\ref{eq:5}) gives
\begin{eqnarray}
T_s&=&\frac{4k^2}{k_B\left(\frac{\partial k}{\partial a}\right)^2}\left\{\frac{2Pv}{a^2}+\frac{1}{6}\frac{\partial^2u_0}{\partial a^2}+\right.\nonumber\\
&+&\left.\frac{\hbar\omega}{4k}\left[\frac{\partial^2k}{\partial a^2}-\frac{1}{2k}\left(\frac{\partial k}{\partial a}\right)^2\right]\right\}.
\label{eq:6}
\end{eqnarray}
Then, the melting temperature $T_m$ is calculated by the Taylor series as \cite{15,16}
\begin{eqnarray}
T_m&\approx&T_s+\frac{a_m-a_s}{k_B\gamma_G(P,T_s)}\left\{\frac{Pv(P,T_s)}{a_s}+\right.\nonumber\\
&+&\left.\frac{1}{18}\left[\left(\frac{\partial u_0}{\partial a}\right)_{T=T_s}+a_s\left(\frac{\partial^2 u_0}{\partial a^2}\right)_{T=T_s}\right]\right\},
\label{eq:7}
\end{eqnarray}
where $a_m=a(P,T_m)$, $a_s=a(P,T_s)$, and $\gamma_G=-\frac{a}{6k}\frac{\partial k}{\partial a}X$ is the Gruneisen parameter. 

In principle, it is possible to access the high-pressure melting boundary via the one-phase approach introduced above. Solving Eq. (\ref{eq:6}) is time-consuming \cite{15}. Moreover, the simplest way to solve Eq. (\ref{eq:7}) is based on an experimental value of $a_m$ \cite{16}. Therefore, we only apply Eq. (\ref{eq:6}) and (\ref{eq:7}) to investigate the melting point at zero pressure. The pressure dependence of $T_m$ can be considered by a dislocation-mediated melting theory \cite{17}, which is
\begin{eqnarray}
T_m(P)=T_m(0)\frac{G(P)}{G(0)}\left(\frac{B_T(P)}{B_T(0)}\right)^{-\frac{1}{b}},
\label{eq:8}
\end{eqnarray}
where $G$ is the shear modulus, $B_T$ is the isothermal bulk modulus, and $b=(dB_T/dP)_{P=0}$. In the framework of the SMM, $G$ and $B_T$ are \cite{33,34}
\begin{eqnarray}
G&=&\frac{k^5}{2\pi a(1+\nu)\left[k^4+\gamma^2\theta^2(X+1)(X+2)\right]},
\label{eq:9}
\end{eqnarray}
\begin{eqnarray}
B_T=-\frac{a(P,T)}{3}\left(\frac{\partial P}{\partial a}\right)_T\left(\frac{a(P,0)}{a(P,T)}\right)^3,
\label{eq:10}
\end{eqnarray}
where $\nu$ is the Poisson ratio. Conventionally, $\nu$ can be interpreted as a slowly varying function of temperature and pressure \cite{17,35}.

In general, the ideal model only requires simple calculations to capture the melting properties of crystals \cite{15,16}. However, the growth rate of $T_m$ sometimes is overestimated \cite{18,22,23}. Thus, in the next subsection, we extend this model to include effects of vacancies on the melting transition.

\subsection{Defective Model}
At the high-temperature regime, some atoms vibrates around their equilibrium positions with large amplitudes and the thermal vibrations can dislodge the atoms from lattice nodes. The dislocation of atoms causes vacancies \cite{19,20,21}. In our theoretical model, a random selection of lattice sites contains a vacancy defect, rather than a particle. Besides, the distance between any two vacancies is supposed to be sufficiently long to ignore their interaction. Based on this assumption and the minimization of the Gibbs free energy, the equilibrium vacancy concentration $n_v$ is provided by (see the Appendix A) \cite{18,22,23}
\begin{eqnarray}
n_\nu=\exp\left(\frac{u_0}{4\theta}\right),
\label{eq:11}
\end{eqnarray}

The temperature is now considered to be a function of both pressure, volume and the equilibrium vacancy concentration. By using the first-order approximation of the Taylor series, the melting temperature of defective crystals $T_m^R$ is written by \cite{18,22,23}
\begin{eqnarray}
T_m^R&=&T_m-\left(\frac{\partial T}{\partial n_\nu}\right)_{P,V}n_\nu(T_m)\nonumber\\
&=&T_m-\frac{T_m^2}{\cfrac{T_m}{4}\cfrac{\partial u_0}{\partial \theta}-\cfrac{u_0}{4k_B}}.
\label{eq:12}
\end{eqnarray}
The vacancy formation breaks down homogeneous atomic bonding and then decreases the mechanical stability limit of solids, so we always have $T_m^R<T_m$ from Eq. (\ref{eq:12}). The reduction in the melting temperature is quantified by a dimensionless quantity $\delta$, which is \cite{18,22,23}
\begin{eqnarray}
\delta=\frac{T_m-T_m^R}{T_m}.
\label{eq:13}
\end{eqnarray}
In previous works \cite{18,22,23}, Eq. (\ref{eq:12}) has been used to investigate the melting transition up to hundreds of GPa. However, one can observe a dramatic growth of $\delta$ with a increase of $T_m$ \cite{18,22,23}. Hence, the high-pressure melting temperature $T_m^R(P)$ may reach a maximum at $P\geq 100$ GPa \cite{18,22,23}. This event is not experimentally expected because the atomic volume and latent heat of fusion vary monotonically with compression \cite{36,37}. Consequently, Eq. (\ref{eq:12}) only gives a good prediction in a pressure range from 0 to 100 GPa \cite{18,22,23} (see the Appendix B).

Recall that the pressure dependence of the melting temperature has an intimate relation with elasticity data (Eq. (\ref{eq:8})) \cite{17}. Notably, in Ref.\cite{38,39}, contributions of vacancies to mechanical responses have been clarified. Specifically, the shear modulus $G^R$ and the isothermal bulk modulus $B_T^R$ of defective crystals are given by 
\begin{eqnarray}
G^R&=&G-\frac{an_\nu}{2\nu(1+\nu)}\left\{\frac{a}{4\theta}\left(\frac{\partial u_0}{\partial a}\right)^2\left(2+\frac{u_0}{4\theta}\right)+\right.\nonumber\\
&+&\left.\left(a\frac{\partial^2u_0}{\partial a^2}+\frac{1}{2}\frac{\partial u_0}{\partial a}\right)\left(1+\frac{u_0}{4\theta}\right)\right\},
\label{eq:14}
\end{eqnarray}
\begin{eqnarray}
B_T^R=B_T+n_\nu n_1\Delta B_T+n_\nu\frac{u_0}{4\psi_0} B_T,
\label{eq:15}
\end{eqnarray}
where $n_1$ is the first coordination number in perfect lattice, $\psi_0$ is the Helmholtz free energy per atom, and $\Delta B_T$ is a vacancy-induced change in the isothermal bulk modulus \cite{38,39}. The obtained results suggest that one can rewrite Eq. (\ref{eq:8}) for defective crystals as 
\begin{eqnarray}
T_m^R(P)=T_m^R(0)\frac{G^R(P)}{G^R(0)}\left(\frac{B_T^R(P)}{B_T^R(0)}\right)^{-\frac{1}{b^R}},
\label{eq:16}
\end{eqnarray}
where $b^R=\left(dB_T^R/dP\right)_{P=0}$ and the input value of $T_m^R(0)$ can be taken from Eq. (\ref{eq:12}). It is important to note that the dislocation-mediated melting theory \cite{17} allows us to determine the right side of Eq. (\ref{eq:16}) at room temperature ($T=300$ K). In addition, Eq. (\ref{eq:11}) reveals that $(\partial n_v/\partial P)_{T=300 K}<0$ when $0< P< P^*$, where the critical pressure $P^*$ obeys $u_0(P^*,300)=\min[u_0]$. Thus, $n_\nu(P^*,300)\ll n_\nu(0,300)\ll n_\nu(0,T_m)=10^{-4}-10^{-2}$ \cite{19,20,21}. The value of $n_\nu(P^*,300)$ is small enough to ignore the difference between the elastic moduli of perfect and imperfect crystals. Hence, we do not need to use the full analytical expression for $n_\nu$ to keep $(\partial n_v/\partial P)_{T=300 K}<0$ in all range of pressures (see the Appendix A). For simplicity, we can use $G^R(P,300)=G(P,300)$ and $B_T^R(P,300)=B_T(P,300)$ when $P\geq P^*$. At the high-pressure regime, the reduction factor $\delta$ calculated by Eqs.(\ref{eq:8}) and (\ref{eq:16}) saturates to $\delta^*$, which is
\begin{eqnarray}
\delta^*=\frac{T_m(P^*)-T_m^R(P^*)}{T_m(P^*)}.
\label{eq:17}
\end{eqnarray}
Consequently, all of the mentioned limitations in Ref.\cite{18,22,23} can be resolved by combining Eqs. (\ref{eq:12}) and (\ref{eq:16}). 
\section{RESULTS AND DISCUSSION}
In this section, first of all, we demonstrate specifically the difference among the ideal model, Eq. (\ref{eq:12}), and the modified defective model by carrying out numerical calculations for Cu, Ag, and Au (subsection III-A). By comparing the obtained results with experiments, simulations, and other calculations, the advantages and limitations of each approach are clarified. Our analysis shows the effectiveness of the defective model in describing the high-pressure melting properties of "real crystals" (crystals with defects in their atomic lattice). Hence, in subsection III-B, we continue to apply this theoretical model to further understand the complex melting tendency of BCC transitions metals including Mo, Ta, and W. 
\subsection{The melting properties of Cu, Ag, and Au}
To describe the interatomic interaction in face-centered cubic (FCC) metals like Cu, Ag, and Au, the SMM analysis typically uses the Mie-Lennard-Jones pair potential as \cite{15}
\begin{eqnarray}
\varphi=\frac{D}{n-m}\left[m\left(\frac{r_0}{r}\right)^n-n\left(\frac{r_0}{r}\right)^m\right],
\label{eq:18}
\end{eqnarray}
where $D$ is the potential well depth, $r_0$ is the equilibrium distance between two atoms, and $m$ and $n$ are adjustable parameters extracted from experiments. The Mie-Lennard-Jones parameters for Cu, Ag, and Au are listed in Table \ref{table:1}.

\begin{table}[h!]
\centering
\begin{tabular}{|c| c| c| c| c|} 
 \hline
 Metal & $D$ (eV) & $m$ & $n$ & $r_0$ (\AA) \\ 
 \hline
 Cu & 0.2929 & 5.5 & 11.0 & 2.5487 \\ 
 Ag & 0.2866 & 5.5 & 11.5 & 2.8760 \\
 Au & 0.4035 & 5.5 & 10.5 & 2.8751\\
 \hline
\end{tabular}
\caption{The Mie-Lennard-Jones parameters for Cu, Ag, and Au \cite{40}.}
\label{table:1}
\end{table}

\begin{figure}[htp]
\includegraphics[width=9 cm]{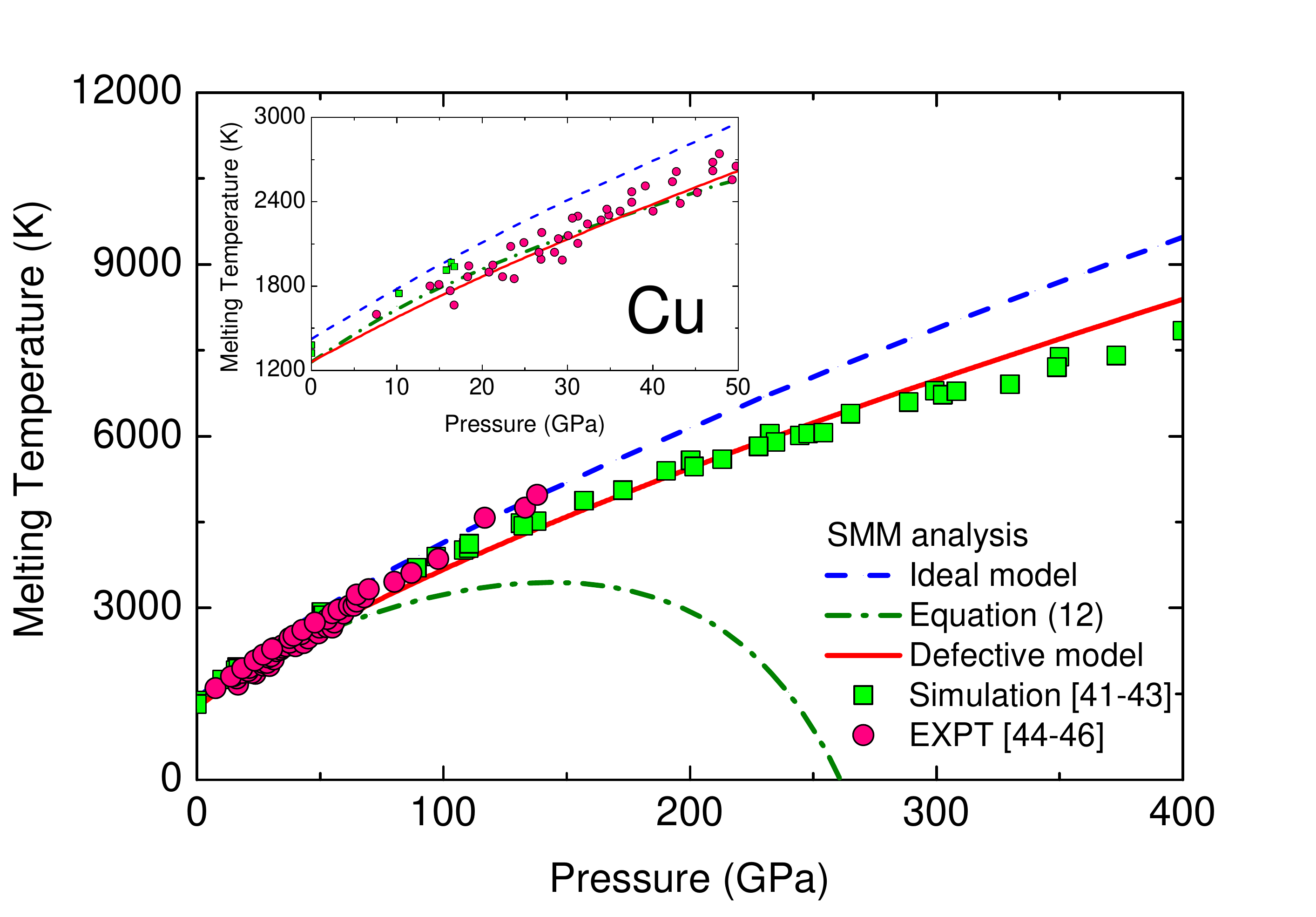}
\caption{\label{fig:1}(Color online) The melting temperature of Cu as a function of pressure determined by the SMM, simulations \cite{41,42,43}, and experiments (EXPT) \cite{44,45,46}. Inset: The same plot as the mainframe but a pressure range from 0 to 50 GPa is zoomed in.}
\end{figure}

Figure \ref{fig:1} shows how the melting temperature of Cu depends on pressure. It is clear to see that the ideal model provides a steep melting curve. The assumption of perfect bonding in the lattice space can cause the superheating of solids above the melting point. Thus, the deviation of our calculations from other simulations \cite{41,42,43} and experimental extrapolates \cite{44,45,46} reaches to 25 $\%$ at 400 GPa. On the other hand, working with vacancies gives us a flatter melting curve \cite{18,22,23}. Physically, the presence of vacancies can trigger the melting transition by creating low atomic density regions with locally high fluctuation amplitudes \cite{21}. Theoretical calculations based on Eq. (\ref{eq:12}) are in good accordance with experiments \cite{45,46} when $0 \leq P \leq 50$ GPa. However, the divergence between $T_m^R$ and $T_m$ leads to a strange melting tendency. The melting slope $dT_m^R/dP$ becomes negative at $P=150$ GPa. According to the Clausius-Clapeyron relation \cite{47,48}, the negative melting slope implies that the density of solid Cu is less than that of liquid Cu at ultra-high pressures. This puzzling conclusion has not been supported by any references \cite{41,42,43,44,45,46}. Notably, our defective model can rapidly resolve all of the mentioned contradictions. The maximum error between our numerical results and existing data \cite{41,42,43,44,45,46} is only 7 \%. This value validates the accuracy of our analytical approach.

\begin{figure}[htp]
\includegraphics[width=9 cm]{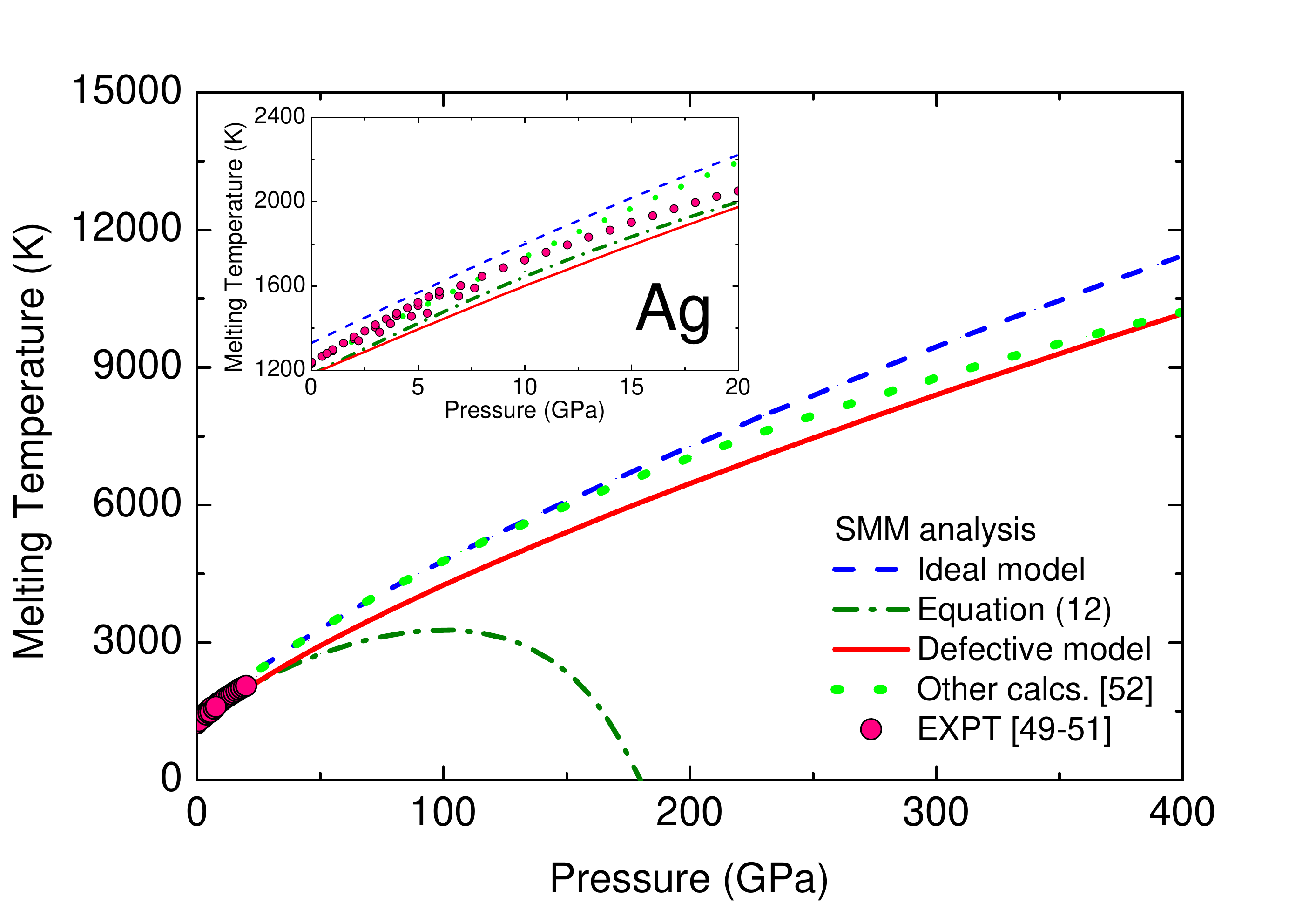}
\caption{\label{fig:2}(Color online) Effects of pressure on the melting temperature of Ag obtained from the SMM, semi-empirical calculations \cite{52}, and experiments \cite{49,50,51}. Inset: Zooming in the low-pressure melting curve of Ag.}
\end{figure}

\begin{figure}[htp]
\includegraphics[width=9 cm]{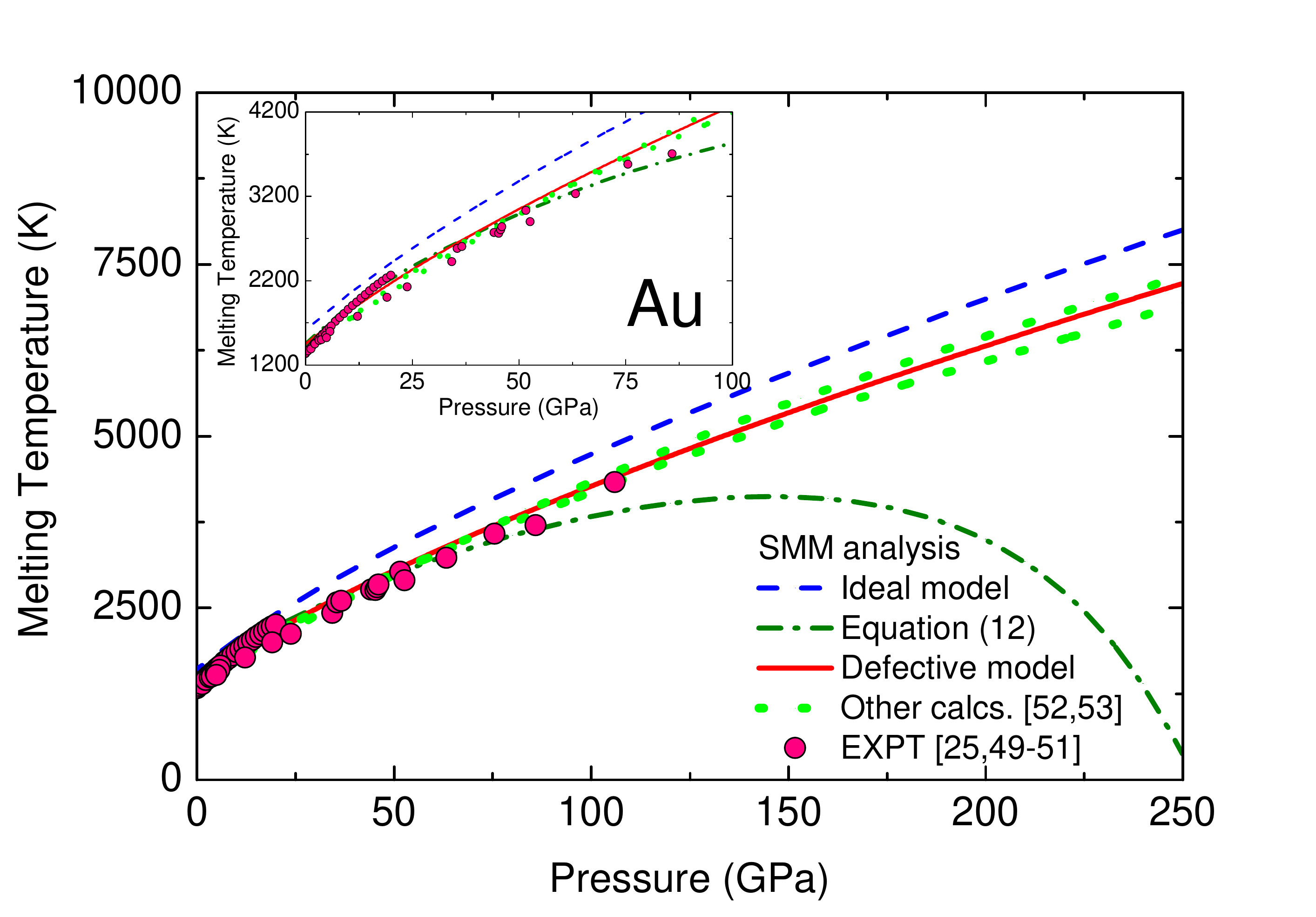}
\caption{\label{fig:3}(Color online) The melting temperature of Au versus pressure given by the SMM, semi-empirical calculations \cite{52}, ab initio \cite{53}, and experiments \cite{25,49,50,51}. Inset: A pressure region $0\leq P\leq 100$ GPa is zoomed in.}
\end{figure}

Figure \ref{fig:2} shows the melting temperature of Ag as a function of pressure. To the best of our knowledge, the melting temperature of Ag is only measured by the differential thermal analysis up to 20 GPa \cite{49,50} and by a Bridgman-type cell up to 8 GPa \cite{51}. In a previous work \cite{52}, Hieu and Ha proposed a semi-empirical model based on Lindemann's criterion and pressure dependence of the Gruneisen parameter to quantitatively explain experimental data and capture the melting behaviors of Ag up to 460 GPa. The melting curves of Ag obtained from the ideal model, the defective model, semi-empirical calculations \cite{52}, and experiments \cite{49,50,51} agree well with each other. Similar to Cu, Eq. (\ref{eq:12}) works effectively at the low-pressure regime, but it fails to predict the melting temperature of Ag under extreme conditions ($P>100$ GPa). Some authors argue that the failure of Eq. (\ref{eq:12}) is due to effects of microstructural transitions \cite{34}. For example, the melting point of Ag at 100 GPa may be close to the FCC-BCC-liquid triple point. However, this theoretical interpretation is ruled out in the case of Au shown in Figure \ref{fig:3}. 

Figure \ref{fig:3} shows the correlations between pressure and the melting temperature of Au up to 250 GPa. It is possible to accurately reproduce experimental results \cite{25,49,50,51} for Au in a pressure range from 0 to 100 GPa by using Eq. (\ref{eq:12}). However, in this case, Eq. (\ref{eq:12}) is invalidated when $P\geq 150$ GPa, while the FCC-BCC transition pressure is reported to be approximately 240 GPa \cite{53}. To understand a monotonous variation of the melting temperature with compression, one can employ the ideal model. Nevertheless, this is not an exact approximation. For example, the ideal model gives $T_m(0)=1612$ K, which is 21 \% larger than the corresponding experimental value of 1335 K \cite{25,49,50,51}. In contrast, combining Eqs. (\ref{eq:12}) and (\ref{eq:16}) provides $\delta^*_{Au}=9.74$ \%. This reduction brings a good agreement between the SMM analysis and comparable data \cite{25,49,50,51,52,53} for Au. Remarkably, growths in the melting temperatures of Cu, Ag, and Au determined by our defective model are similar to each other. This conclusion is consistent with a $d$-electron band mediated melting theory \cite{54}. Cu, Ag, and Au have the full-filled $d$ electron band, so their melting curves should have the same form \cite{54}. 

\subsection{The melting properties of Mo, Ta, and W}
For BCC metals such as Mo, Ta, and W, the Morse potential has widely been employed to describe interatom interactions \cite{55,56}
\begin{eqnarray}
\varphi=D^*\left[e^{-2\alpha(r-r_0^*)}-2e^{-\alpha(r-r_0^*)}\right],
\label{eq:19}
\end{eqnarray}
where $D^*$ is the dissociation energy, $\alpha$ is a constant, and $r_0^*$ is an equilibrium value obeying $\varphi(r_0^*)=-D^*$. The Morse parameters for Mo, Ta, and W are presented in Table \ref{table:2}. 

\begin{table}[h!]
\centering
\begin{tabular}{|c| c| c| c|} 
 \hline
 Metal & $D^*$ (eV) & $\alpha$ (\AA$^{-1}$) & $r_0^*$ (\AA) \\ 
 \hline
 Mo  & 0.8032 & 1.5079 & 2.9760 \\ 
 Ta  & 0.8761 & 1.3139 & 3.0580 \\
 W   & 0.8900 & 1.4400 & 3.0520 \\
 \hline
\end{tabular}
\caption{The Morse parameters for Mo, Ta, and W \cite{57,58}.}
\label{table:2}
\end{table}

\begin{figure}[htp]
\includegraphics[width=9 cm]{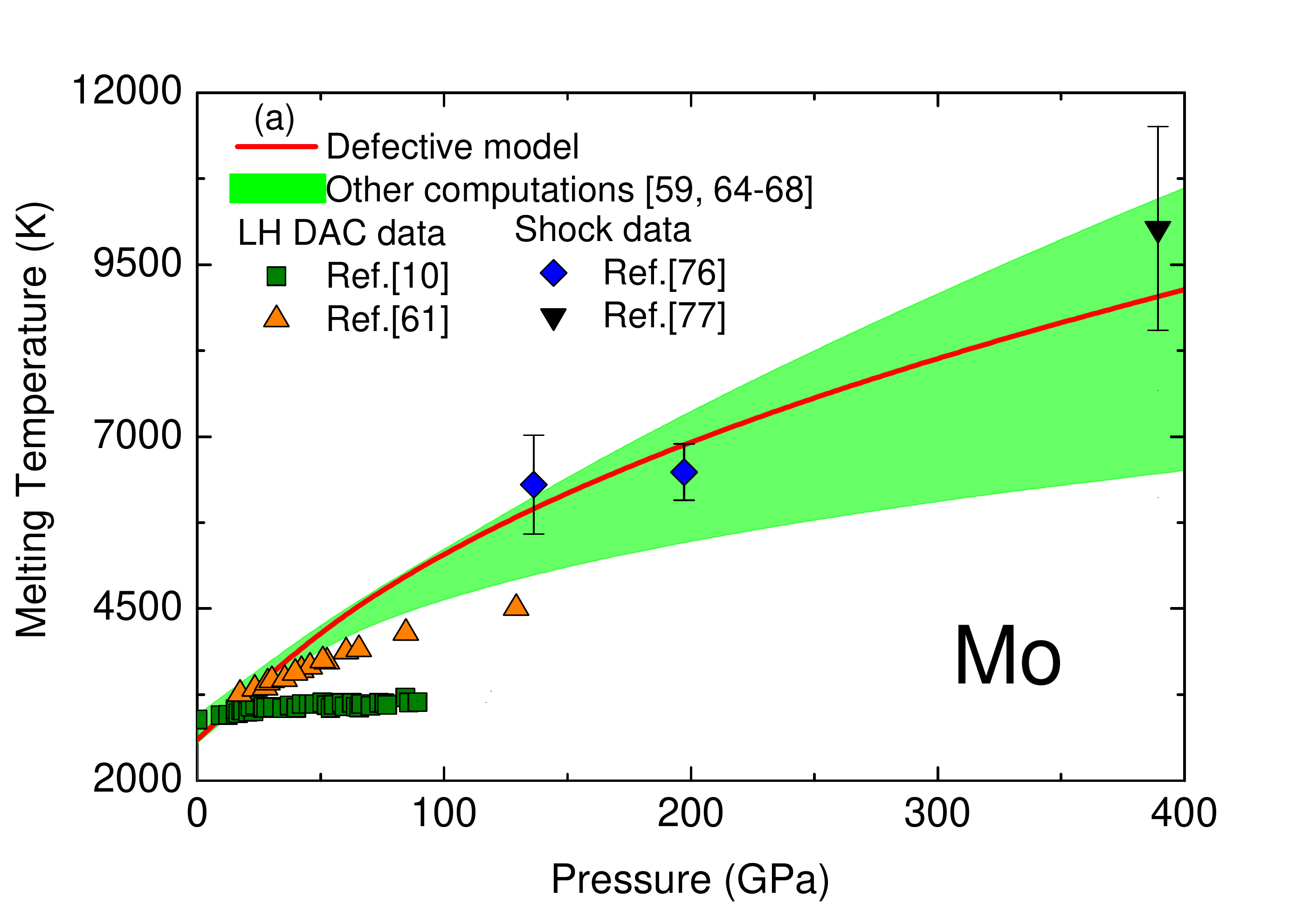}
\includegraphics[width=9 cm]{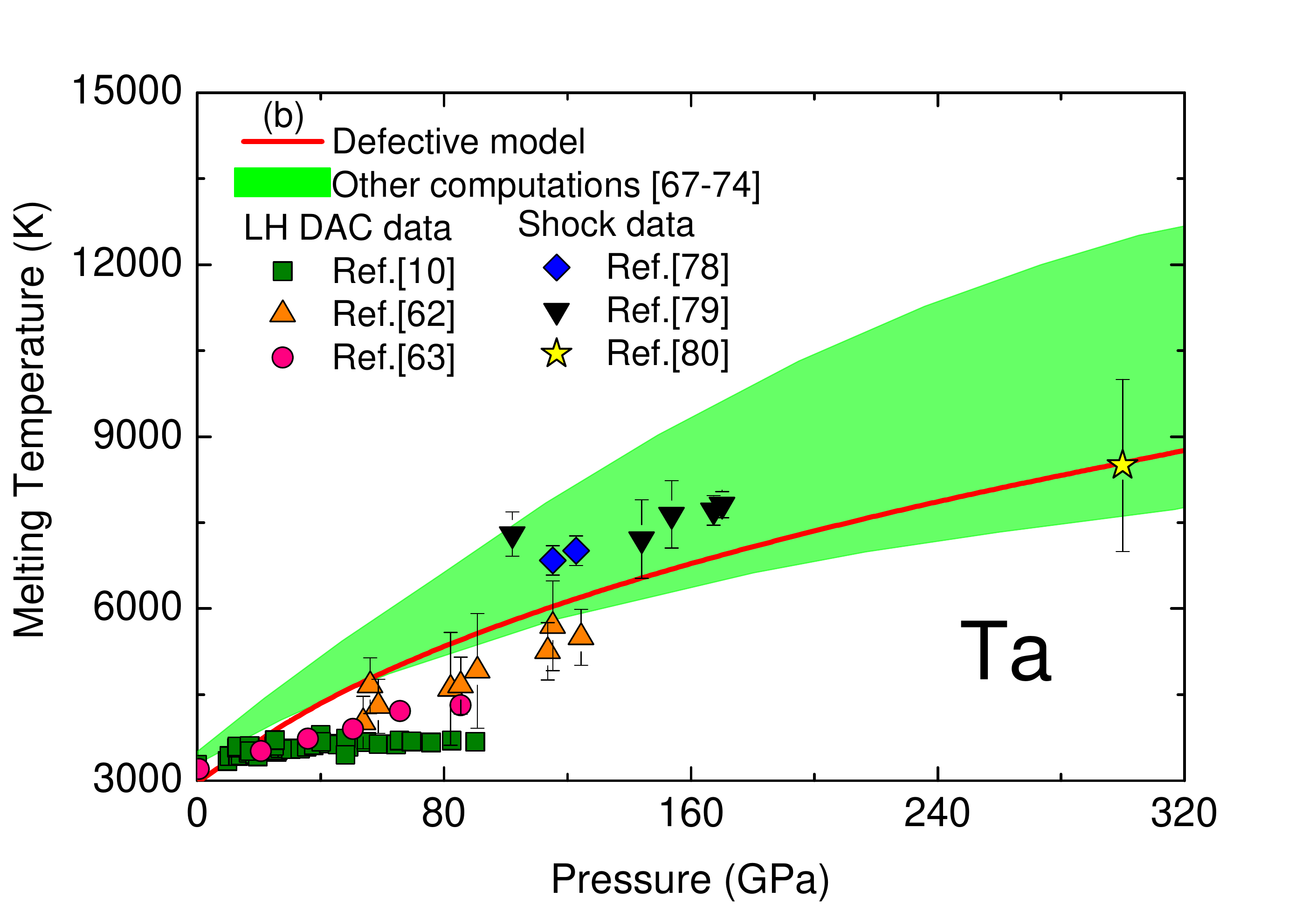}
\includegraphics[width=9 cm]{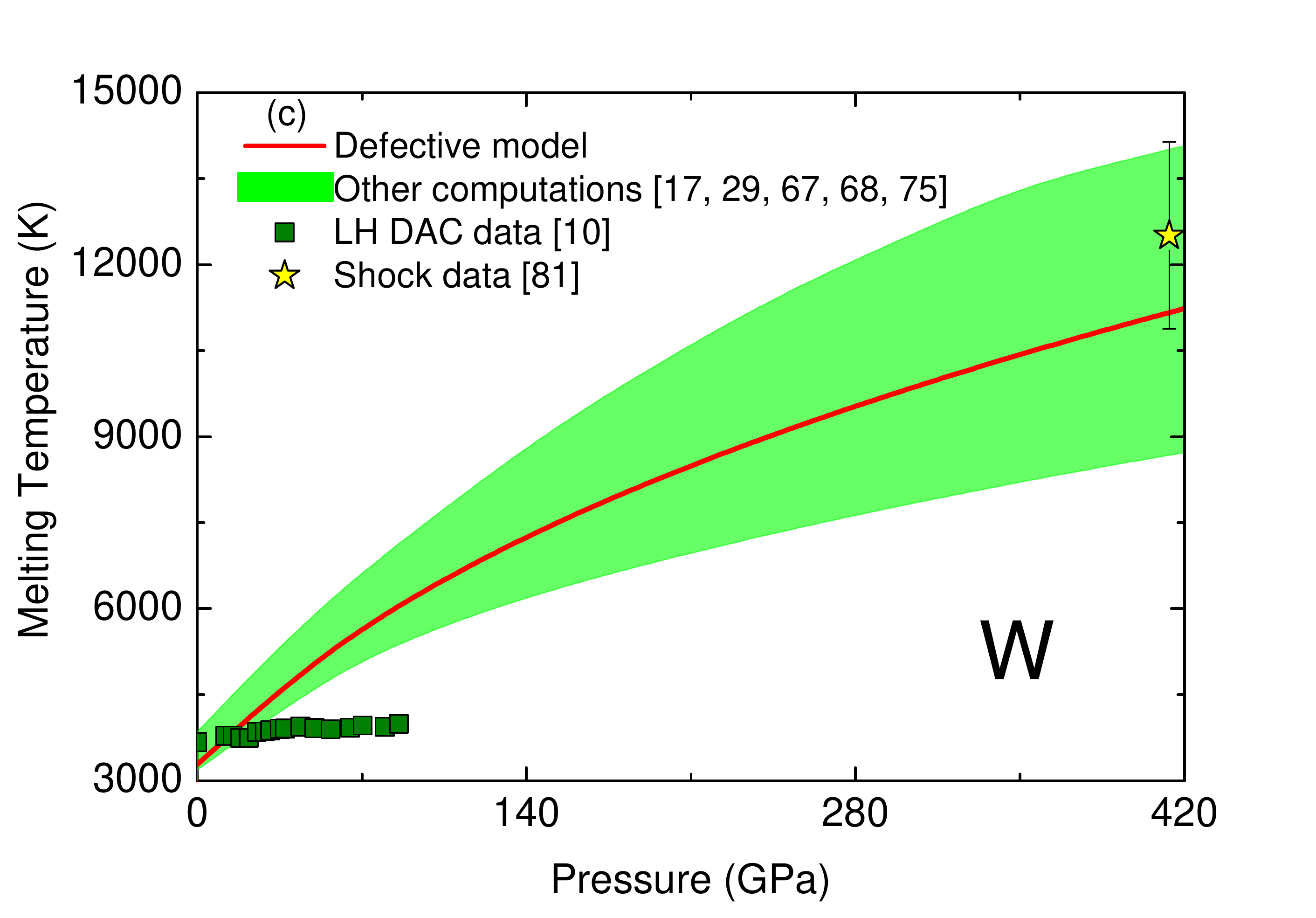}
\caption{\label{fig:4}(Color online) The high-pressure melting curve of (a) Mo, (b) Ta, and (c) W given by the SMM and the existing data \cite{10,17,29,59,61,62,63,64,65,66,67,68,69,70,71,72,73,74,75,76,77,78,79,80,81}.}
\end{figure}

Figure \ref{fig:4} shows effects of pressure on the melting behaviors of Mo (Figure 4a), Ta (Figure 4b), and W (Figure 4c). One can realize two predicted regions for the melting temperature of Mo, Ta, and W including

(i) \textit{LH DAC data region}: In Ref.\cite{10}, the average melting slope of Mo, Ta, and W has been reported to be approximately zero. This melting tendency has become a debated topic for decades. Theoretically, Benoloshko and co-workers \cite{59} conceive that there was a confusion between solid-solid and solid-liquid transitions in previous LH DAC experiments \cite{10}. The observed flow event is not caused by melting but caused by internal non-hydrostatic stresses \cite{59}. They have simulated the FCC-BCC boundary of Mo and shown a good accordance with the flat melting curve given by LH DAC \cite{59}. However, the existence of the FCC phase in Ref.\cite{59} is anomalous. Instead, the BCC Mo lattice can be stabilized by transforming into the hexagonal close-packed structure \cite{60}. Furthermore, the assumption of Belonoshko and co-workers \cite{59} is excluded due to the detection of chemical contaminations in LH DAC measurements \cite{61,62,63}. 

In practice, the melting point of the laser-heated sample can become anomalously low due to carbidation processes and chemical activities with the pressure-transmitting media \cite{61,62,63}. Additionally, experimental observations are also perturbed when the pressure medium melts \cite{61,62,63}. Experimentalists have tried for almost twenty years to develop LH DAC techniques and achieve the high-slope melting curve of Mo, Ta, and W. Unfortunately, the obtained results have been still inconsistent and have not truly agreed with theoretical predictions and dynamic shock-wave measurements \cite{61,62,63}. Consequently, more efforts are needed to overcome these challenges.  

(ii) \textit{Theoretical and shock data region}: In contrast to LH DAC data, theoretical computations \cite{17,29,59,64,65,66,67,68,69,70,71,72,73,74,75} and dynamic shock-wave methods \cite{76,77,78,79,80,81} predict significant growth in the melting temperature of the compressed Mo, Ta, and W. The SMM calculations based on the defective model are consistent with this melting tendency. A large value of $\delta^*$ ($\approx10$ $\%$) for Mo, Ta, and W proves the crucial roles of vacancies on the melting mechanism. Notably, our numerical results for Ta (shown in Figure 4b) satisfy not only theoretical expectations \cite{67,68,69,70,71,72,73,78,79,80} but also recent LH DAC experiments \cite{62}. Thus, the real melting curve of Ta may be very close to our estimations. Our numerical data can be quantitatively described by the Simon relation as \cite{82}
\begin{eqnarray}
T_m=T_{m0}\left(\frac{P}{P_0}+1\right)^c,
\label{eq:20}
\end{eqnarray}
where fitting parameters $T_{m0}$, $P_0$, and $c$ are presented in Table \ref{table:3}. If we can take into account the electronic structure and construct a new interatomic potential based on the typical properties of both solids and liquids, the SMM analysis is expected to be more accurate. 

\begin{table}[h!]
\centering
\begin{tabular}{|c| c| c| c| c|} 
 \hline
 Metal & $T_{m0}$ (K) & $P_0$ (GPa) & $c$ & $\delta^*$ (\%) \\ 
 \hline
 Cu & 1283.2965 & 27.2348 & 0.6815 & 11.45\\ 
 Ag & 1215.1917 & 19.8905 & 0.6962 & 11.13\\
 Au  & 1468.7550 & 24.9996 & 0.6639 & 9.74\\
 Mo  & 2588.3560 & 23.8435 & 0.4391 & 10.17\\
 Ta  & 2956.6524 & 24.4968 & 0.4111 & 8.62\\
 W  & 3276.9478 & 28.0036 & 0.4450 & 10.92\\
 \hline
\end{tabular}
\caption{The Simon parameters \cite{82} and the critical reduction factor $\delta^*$ for Cu, Ag, Au, Mo, Ta, and W extracted from our defective model.}
\label{table:3}
\end{table}

\section{CONCLUSION}
We have combined the SMM analysis with the dislocation-mediated melting theory to investigate effects of vacancies on the melting properties of crystals. Numerical calculations have been applied to predict the melting temperature of Cu, Ag, Au, Mo, Ta, and W up to 400 GPa. The obtained results have proven that the melting transition is strongly facilitated by the presence of vacancies. Our theoretical calculations are in quantitative accordance with previous experiments, simulations, and other models. Consequently, our simple analytical approach would improve understanding of the melting mechanisms in crystalline materials under extreme conditions. It is possible to expand our defective model to investigate the melting phenomenon in low-dimensional or multicomponent systems. 

\appendix
\section{The Equilibrium Vacancy Concentration}
According to statistical dynamics, the key quantity determining a proliferation of vacancies is the Gibbs free energy \cite{18,19,20,21,22,23}. Fundamentally, one can directly link the Gibbs free energy of a defective monatomic crystal, $g^R(P,T,n)$, to its perfect counterpart, $g(P,T)$, by \cite{83,84}
\begin{eqnarray}
g^R(P,T,n)=g(P,T)+ng_{\nu}^{f}(P,T)-TS_c(n),
\label{eq:A1}
\end{eqnarray}
where $g_{\nu}^f$ is the Gibbs free energy cost of forming a single thermal vacancy, and $S_c=k_B\ln\frac{(N+n)!}{N!n!}$ is the configurational entropy of the binary mixture made by $N$ atoms and $n$ vacancies. Based on the Stirling approximation, $S_c$ can be rewritten by
\begin{eqnarray}
S_c&=&k_B(N+n)[\ln(N+n)-1]-\nonumber\\
&-&k_BN(\ln N-1)-k_Bn(\ln n-1).
\label{eq:A2}
\end{eqnarray}
Since vacancies exist in thermodynamic equilibrium, $g^R(P,T,n)$ satisfies the minimum condition as \cite{18,19,20,21,22,23}
\begin{eqnarray}
\left(\frac{\partial{g^R}}{\partial{n}}\right)_{P,T}=0.
\label{eq:A3}
\end{eqnarray}
Combining Eqs.(\ref{eq:A1}), (\ref{eq:A2}) and (\ref{eq:A3}) gives
\begin{eqnarray}
n_{\nu}=\frac{n}{N+n}=\exp\left(-\frac{g_{\nu}^{f}}{\theta}\right).
\label{eq:A4}
\end{eqnarray}

In our theoretical model, when the 0-th atom escapes from its lattice site and moves to free positions on the surface of the crystal, the movement creates a vacancy. The total change in the Helmholtz free energy of the 0-th atom and its new neighbors is assumed to be $(B-1)\psi_0$, where $B$ is a scalar factor \cite{83,84}. After diffusing to the surface, the 0-th atom can restore more than half of the broken bonds \cite{85}, so one has $B>1$ \cite{83,84}. Besides, since the presence of vacancy reduces the bulk coordination number, the Helmholtz free energy of the $i$-th atom also increases from $\psi_0$ to $\psi_i$. Consequently, an analytical expression of $g_\nu^f$ at zero pressure is
\begin{eqnarray}
g_\nu^f&=&(B-1)\psi_0+\sum_i(\psi_i-\psi_0).
\label{eq:A5}
\end{eqnarray}
The harmonic forms of $\psi_0$ and $\psi_i$ are \cite{83,84}
\begin{eqnarray}
\psi_0=\frac{1}{2}u_0+3\theta[x+\ln(1-e^{-2x})],
\label{eq:A6}
\end{eqnarray}
\begin{eqnarray}
\psi_i=\frac{1}{2}(u_0-\varphi_{i0})+3\theta[x+\ln(1-e^{-2x})].
\label{eq:A7}
\end{eqnarray}
Inserting Eqs.(\ref{eq:A6}) and (\ref{eq:A7}) into Eq.(\ref{eq:A5}) provides
\begin{eqnarray}
g_\nu^f=-\frac{u_0}{2}+(B-1)\psi_0.
\label{eq:A8}
\end{eqnarray}
Because $g_\nu^f>0$ \cite{83,84}, the scale factor $B$ in Eq.(\ref{eq:A8}) is limited by
\begin{eqnarray}
1<B<1+\frac{u_0}{2\psi_0}.
\label{eq:A9}
\end{eqnarray}
In the present work, we take the average value of $B$ as
\begin{eqnarray}
B\approx1+\frac{u_0}{4\psi_0}.
\label{eq:A10}
\end{eqnarray}
Applying Eqs.(\ref{eq:A4}), (\ref{eq:A8}) and (\ref{eq:A10}) gives
\begin{eqnarray}
n_v=\exp\left(\frac{u_0}{4\theta}\right).
\label{eq:A11}
\end{eqnarray}

\begin{figure}[htp]
\includegraphics[width=9 cm]{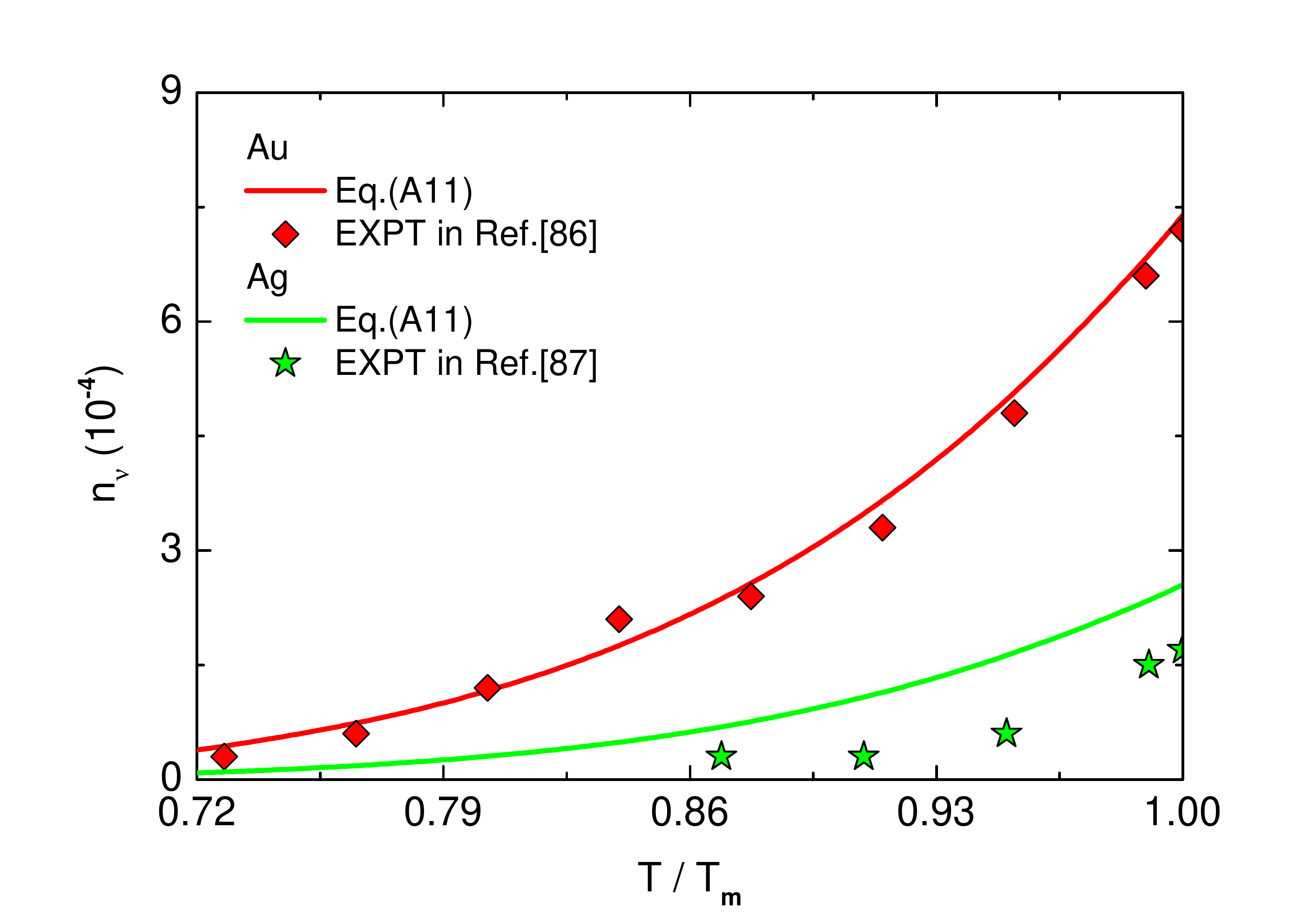}
\caption{\label{fig:5} (Color online) The equilibrium vacancy concentration of Au and Ag as a function of normalized temperature at zero pressure determined by the SMM and EXPT \cite{86,87}.}
\end{figure}

Figure \ref{fig:5} and Table \ref{table:4} show how vacancies are thermally generated in Cu, Au, Ag, Mo, Ta, and W at zero pressure. A good agreement between numerical calculations and experimental data \cite{19,86,87,88} validates our chosen value of $B$. One can derive a general equation for determining the equilibrium vacancy concentration in the framework of the SMM as
\begin{eqnarray}
n_v=\exp\left(\frac{u_0}{4\theta}\right)\exp\left(\frac{-P\Delta v_f}{\theta}\right),
\label{eq:A12}
\end{eqnarray}
where $\Delta v_f$ is the vacancy formation volume. Calculating $\Delta v_f$ requires sophisticated techniques. The correlation between the melting temperature and the equilibrium vacancy concentration based on Eq.(\ref{eq:A12}) is very complicated. Hence, in the previous works \cite{18,22,23}, authors only used Eq.(\ref{eq:A11}) to capture vacancy effects on the melting transition. Table \ref{table:4} reveals a reasonable reason for considering the low-pressure melting curve of defective crystals via Eqs. (\ref{eq:A11}) and (\ref{eq:12}).

\begin{table}[h!]
\centering
\begin{tabular}{|c| c| c| c| c| c| c|} 
 \hline
 Metal & Cu & Ag & Au & Mo & Ta & W \\ 
 \hline
SMM & 3.8 & 2.6 & 7.4 & 0.38 & 4.0\footnote{Taken from Ref.\cite{23}} & 1.5 \\
EXPT \cite{19} & 2-7.6 & 1.7-5.2 & 7 & 0.43\footnote{Taken from Ref.\cite{88}} & $3.5^b$ & 1-3\\
 \hline
\end{tabular}
\caption{$n_\nu$ ($10^{-4}$) of Cu, Ag, Au, Mo, Ta, and W at $T_m(0)$ obtained from the SMM and EXPT \cite{19,88}.}
\label{table:4}
\end{table}

\section{The Failure of the Previous SMM Defective Model at High Pressures}
In Ref.\cite{18,22,23}, the number of vacancies at the melting point plays a crucial role in investigating the defect-mediated melting mechanism. Phenomenologically, the generation of thermal vacancies causes the bulk expansion during the melting process \cite{89,90}. This physical perspective allows us to adopt \cite{89,90}
\begin{eqnarray}
\Delta v_f=\Delta v_m\frac{\Delta H_f}{\Delta H_m},
\label{eq:B1}
\end{eqnarray}
where $\Delta v_m$ is the volume change on fusion, $\Delta H_m$ is the latent heat of fusion, and $\Delta H_f$ is the vacancy formation enthalpy. By collecting experimental data for 35 elements at zero pressure, Bollman found $\Delta H_f/\Delta H_m\approx8$ \cite{91}. Notably, Errandonea and co-workers \cite{11} have successfully reproduced the previous LH DAC data \cite{10} for Ta via this simple relation and the first-principle calculations for $\Delta H_f$ \cite{92}. Thus, we can rewrite Eq.(\ref{eq:B1}) by  
\begin{eqnarray}
\Delta v_f=8\Delta v_m
\label{eq:B2}
\end{eqnarray}

From Eqs.(\ref{eq:A12}) and (\ref{eq:B2}), the equilibrium vacancy concentration along the solid-liquid boundary is given by  
\begin{eqnarray}
n_\nu=\exp\left(\frac{u_0}{4\theta}\right)\exp\left(\frac{-8P\Delta v_m}{\theta}\right).
\label{eq:B3}
\end{eqnarray}
When $100<P<400$ GPa, several simulation studies \cite{75,93} suggest that the $\exp(-8P\Delta v_m/\theta)$ term in Eq.(\ref{eq:B3}) remains nearly unchanged and insensitive to compressions. Hence, Eq.(\ref{eq:A11}) can qualitatively describe a proliferation of vacancies along the high-pressure melting curve. However, the equilibrium vacancy concentration obtained from Eq.(\ref{eq:A11}) is overestimated. By using simulation data, one has $\exp(-8P\Delta v_m/\theta)\approx0.064$ for W \cite{75} and $\exp(-8P\Delta v_m/\theta)\approx0.036$ for Cu \cite{93}. Consequently, the Taylor expansion in Eq.(\ref{eq:12}) is invalidated and the previous SMM defective model \cite{18,22,23} fails to predict the melting behaviors of crystals under extreme conditions.

\end{document}